# Superconducting gap structure and bosonic mode in $La_2PrNi_2O_7$ thin films at ambient pressure

Shengtai Fan[1†], Mengjun Ou[1†], Marius Scholten[2], Qing Li[1], Zhiyuan Shang[1], Yi Wang[1], Jiasen Xu[1], Huan Yang[1*], Ilya M. Eremin[2*], Hai-Hu Wen[1*]

[1] National Laboratory of Solid State Microstructures and Department of Physics, Jiangsu Physical Science Research Center, Collaborative Innovation Center of Advanced Microstructures, Nanjing University, Nanjing 210093, China.

[2] Institut für Theoretische Physik III, Ruhr-Universität Bochum, D-44801 Bochum, Germany

† These authors contribute equally to the work.

*Corresponding authors: huanyang@nju.edu.cn, Ilya.Eremin@ruhr-uni-bochum.de, hhwen@nju.edu.cn

**The recent discovery of high temperature superconductivity in nickelate systems has generated tremendous interests in the field of superconductivity. The core issue to understand the superconductivity mechanism is about the superconducting gap and its symmetry. By using the substrate of $SrLaAlO_4$ (001), we have successfully synthesized the superconducting thin film of $La_2PrNi_2O_7$ with $T_c^{onset}$ = 41.5 K. Superconducting tunneling spectra are successfully measured on the terraces after we expose the superconducting layer by using the tip-excavation technique. The spectrum shows a two-gap structure with $\Delta_1 \approx 19$ meV, $\Delta_2 \approx$ 6-8 meV, and fittings based on the Dynes model indicate that the dominant gap should have an anisotropic *s*-wave structure, this allows us to put the priority in selecting the $s^{\pm}$ among the two arguable**

**pairing models: $s^{\pm}$ and *d*-wave. Furthermore, a clear bosonic mode with energy $\Omega \approx 30\pm2$ meV is observed, which further supports a sign reversal gap. Our results shed new light in understanding the mystery of superconductivity in bilayer nickelate superconductors.**

## Main

Superconductivity in nickelates opens up a new era in the research of unconventional high-temperature superconductivity. The starting point was from the observation of the superconductivity in the infinite-layer nickel-based $Nd_{1-x}Sr_xNiO_2$ and its sister compounds[1-2] (for a review see Ref.3). Later on, the breakthrough in nickelate superconductors was the observation of high-temperature superconductivity with the onset transition temperature ($T_c^{onset}$) up to 80 K in bulk $La_3Ni_2O_7$ samples under the pressure above 14 GPa[4,5]. Superconductivity under pressure has also been observed in the Ruddlesden-Popper bilayer $La_2PrNi_2O_7$ sample[6] and $La_2SmNi_2O_{7-\delta}$ sample[7], the trilayer $La_4Ni_3O_{10-\delta}$ samples[8-10], as well as the $La_5Ni_3O_{11}$ sample[11]. Recently, superconductivity with $T_c^{onset}$ of about 40 K was observed at ambient pressure in the $La_3Ni_2O_7$ thin films[12], $(La,Pr)_3Ni_2O_7$ thin films[13], and $La_{3-x}Sr_xNi_2O_7$ thin films[14] with an in-plane compressive strain, which opens a new chapter of the research on high-temperature superconductivity in nickelates at ambient pressure.

The observation of superconductivity in nickelate superconductors inspires scientists to investigate the electronic structure and the superconducting mechanism, especially in the Ruddlesden-Popper (RP) 327 system that holds the high-$T_c$ superconductivity. The angle-resolved photoemission spectroscopic (ARPES) measurements reveal the band structure in $La_3Ni_2O_7$ at ambient pressure. This consists of the α and β pockets from the $3d_{x^2-y^2}$ orbital with sizeable hybridization of the $3d_{z^2}$ orbital which is responsible for the unusually large bonding-antibonding splitting between them. In addition, a flat bonding γ-band is derived from the $3d_{z^2}$ orbital which may exceed or lie just below the Fermi level[15,16]. Under high pressure, the bonding γ band may cross the Fermi energy, which was argued to be closely related to the high-$T_c$ superconductivity[4,17-31]. This is different from cuprates in which the dominant contribution to superconductivity comes from the $3d_{x^2-y^2}$ band. However, it remains controversial whether the γ pocket is absent[32] or blurred[33] in the 327 compressively strained thin films, which makes the origin of high-$T_c$ superconductivity more complex and interesting. According to the varieties of theoretical models, they may be categorized into two major camps for the superconducting pairing symmetry for the bilayer pressurized nickelate 327 bulks. The first one supports the so-called $s^{\pm}$-wave[19,20,27-31] scenario in which the Cooper-pairing appears as a result of the dominant interlayer interaction with dominant interlayer gap $\Delta_{\perp}$, which is promoted by the strong hybridization of $3d_{z^2}$ orbitals between the layers and a smaller intralayer component $\Delta_{||}$ which results from the intralayer interaction. The gap is then positive, $\Delta_{\perp}+\Delta_{||}$ on the symmetric (bonding) bands (which are $\alpha$ and $\gamma$-bands) and negative $\Delta_{||}-\Delta_{\perp}$ on the antisymmetric (antibonding) $\beta$ band, see for example (Ref.19 and Ref.34). In the second scenario the dominant in-plane repulsive interaction due to antiferromagnetic spin fluctuations drives the *d*-wave symmetry of the superconducting state[28,29,35-37], which is somewhat similar to the cuprates. However, the experimental data supporting any gap symmetry is rare because of the very high pressure to achieve superconductivity, besides a preliminary study by using the Andreev reflection under pressure in $La_3Ni_2O_7$[38] and more recently ARPES experiments in thin flims[39]. Therefore, high-quality experimental data for showing the gap structure and intimately pinning down the superconducting mechanism are highly desired. Thanks to the discovery of superconductivity in 327 films with high-$T_c$ superconductivity at ambient pressure, it is now possible to measure the gap structure by using the commonly used tools, such as the scanning tunneling microscopy and spectroscopy (STM/STS).

In this work, we report the successful growth of the $La_2PrNi_2O_7$ thin films with $T_c^{onset}$ up to 41.5 K. We also successfully carried out the single particle tunneling measurements on the film, and obtained the tunneling spectra in the superconducting region constructed by the tip-excavation technique. Although the global resistive transition of the film is still broad, since the STM measurement can allow us to measure the local regions with good superconductivity, we can thus carry out valid measurements of the superconducting spectra with the important message of the gap structure. Repeated measurements show clearly the gap structure and magnitude, and even more for a bosonic mode. The fitting results strongly suggest the dominance of anisotropic *s*-wave gap(s) in the film, supporting the Cooper-pairing scenario of $s^{\pm}$. These results provide key information in understanding the superconducting mechanism in superconducting $La_2PrNi_2O_7$ thin films.

## Superconductivity in $La_2PrNi_2O_7$ films

The 2-unit-cell (2UC) $La_2PrNi_2O_7$ thin films were grown by pulsed laser deposition on (001)-oriented $SrLaAlO_4$ substrates with an in-situ ozone annealing process (see Methods for details). Temperature dependent resistivity of the $La_2PrNi_2O_7$ thin film shows good metallic behavior and a clear superconducting transition (Fig. 1a). $T_c^{onset}$ defined by the cross point of the red lines in the inset of Fig. 1a is about 41.5 K which is comparable to the previously reported values of $T_c$ on bilayer compressively strained nickelate thin films[12-14]. The value of resistivity is close to zero at 2 K, showing a relatively broad transition. This broad resistive transition may be induced by the inhomogeneity of the superconducting phase, especially for these extremely thin and two-dimensional films. However, we want to emphasize again that this does not prevent us from measuring valid tunneling spectra concerning superconductivity since STM is a very local tool, one can always find some local regions with good superconductivity. Figure 1b shows the magneto-transport measurements on a $La_2PrNi_2O_7$ thin film with applied magnetic field along the *c*-axis, and the superconducting transition is gradually suppressed under magnetic fields. Figure 1c shows the temperature dependent upper critical fields ($\mu_0H_{c2}$) with a criterion of 90% of normal-state resistivity $\rho_n(T)$. The data can be well fitted by Ginzburg-Landau based formula, and the fitting yields $\mu_0H_{c2}$ at zero temperature of about 115 T.

Figure 1d shows the X-ray diffraction (XRD) 2θ-ω scans of our $La_2PrNi_2O_7$ thin film and the $SrLaAlO_4$ substrate. All the (00*l*) peaks can be well indexed by the bilayer structure, and there is no signature of impurity phases. The tiny peaks marked with asterisks are found to be induced by the substrate since they appear even with the bare substrate (shown at the bottom of Fig. 1d). The *c*-axis lattice constant calculated from the XRD data is about 20.78 Å which is about 1% elongated compared to bulk sample[6] and in good agreement with the previous reported thin films[13]. The surface morphological features of the thin films are checked by atomic force microscopy at room temperature (Fig. 1e) and STM measurements at low temperatures (Fig. 1f), and the root-mean-square roughness is about 220 pm.

## Observation of superconducting gap and bosonic mode

The surfaces of the thin films are rather flat; however, they are somewhat degenerated and may also be contaminated when exposed to air, thus they show some kind of disorders on the nanometer scale. An example is shown in Fig. 1f, where one can see that the surface shows a flat background with some grains. The calculated root-mean-square roughness is about 0.17 nm, which is a challenge for the measurements of atomically resolved scanning tunneling spectra. In order to get a fresh surface with superconductivity, we use the tip-excavation technology[40], i.e., applying a bias pulse of 2-4 V to the tip when it is stabilized in the initial setpoint conditions of $V_{set}$ = 8 V and $I_{set}$ = 10-20 pA. The pulse may cause an abrupt change in the distance between the tip and the film and take away some part of the film. The topographic images before and after the excavation treatment are shown in Extended Data Fig. 1a and 1b, or in Extended Data Fig. 2a and 2b for another area. One can see that a large part of the covered film has been removed by the tip-excavation process. Although we have not got an atomically resolved surface, this does not rule out the possibility of measuring the single particle tunneling spectrum since the STM is a tool for local measurements. Luckily, in the case reported in the main text, we obtained a terrace structure with two layers of films on the substrate, and the height differences between the two neighboring layers are both about 2 nm, just 1UC. Actually, successful tunneling measurements for superconducting spectroscopy have been achieved even with a scanning on the surface Layer 1 at a certain distance. The nature of the substrate is confirmed by the tunneling spectra which show a very large insulating gap (Extended Data Fig. 1c).

As said, the top layer may have been contaminated in the air during the sample transferring process, no good tunneling spectra can be directly measured on the surface of the films although the gapped feature can still be seen sometimes, but the coherence peaks are not clear, see for example for another area after tip-excavation (Extended Data Fig. 2f). Therefore, we carried out the tunneling spectrum measurements on the inner layer (Layer 1) of the film which is exposed by the tip-excavation technique. The extremely nice point is that we achieved a terrace structure with the planes corresponding to each surface of the two unit-cells. One can see clearly the steps from substrate to Layer 1 and then to Layer 2, as shown in Fig. 2a and 2b. A set of tunneling spectra measured along the arrowed line on Layer 1 and at 0.4 K is shown in Fig. 2c. All the spectra show clear superconducting shapes with coherence-peak features appearing at ± (20±0.6) meV. Meanwhile, there are some kink structures appearing at about ± 6 ~ ± 8 meV, and such features are clearer in the second derivative curve $d^2I/dV^2$ versus bias voltage (Fig. 2d). This kink structures, which can be seen on all the spectra measured on Layer 1, can be related to another smaller superconducting gap. We should emphasize that the two-gap feature has also been observed on the inner surface in another region of the film (Extended Data Fig. 2 and 3), and it is a common feature in the $La_2PrNi_2O_7$ thin films. Besides the coherence peaks at about ± 20 meV, one can see another peak at very high energies of about ± 49 meV which are usually regarded as the bosonic mode. Interestingly, spectra with very similar shape can be obtained in the same region when following another trajectory, as shown in Extended Data Fig.4. The mode energy $\Omega$ can be determined by the middle of the right (or left) wing of the peak at positive (or negative) energies at $\pm(\Delta+\Omega)$[41-43]; meanwhile, it can also be determined by the local dip (or peak) position on the $d^2I/dV^2$ vs. V curves. From Figs. 2c and 2d, we can derive the value of the characteristic energy $\Delta+\Omega$ of about 49±2 meV. This is the first time to illustrate the superconducting spectra as well as the bosonic mode in bilayer nickelate superconducting systems.

## Temperature dependent evolution of the superconducting spectrum

In order to confirm that the gapped feature on the tunneling spectra corresponds to superconductivity, we carry out the measurements of spectra at different temperatures (Fig. 3a). With the increase of temperature, the coherence peak feature at about ± 20 meV is suppressed while the differential conductance at zero bias is lifted. The gapped feature almost disappears on the spectrum measured at about 40 K, being close to $T_c^{onset}$. Therefore, we believe the gapped feature surely comes from superconductivity. Meanwhile, the mode feature disappears simultaneously with the coherence peaks, suggesting its close relationship with superconductivity. We use the spectrum measured at 40 K as the background and normalize the spectra measured at other temperatures, and the result is shown in Fig. 3b.

The spectra measured at low temperatures show a generally gapped feature below about 20 meV; meanwhile, the spectra exhibit somewhat like a “U” shape and a kink appears at low energies. There is also a significant conductance at zero bias. Usually, a V-shaped spectrum near zero bias is a signature of a nodal gap. For example, the observation of a V-shaped spectrum in $Nd_{1-x}Sr_xNiO_2$ thin films indicates a dominant nodal *d*-wave gap in the material from our previous work[40]. In order to get a comprehensive understanding of the gap structure, we carry out careful fitting based on the Dynes model[44]. Although the gap energies are symmetric and spatially rather uniform (Fig. 2c and d, and Extended Data Fig. 4), the spectrum shows a clear particle-hole asymmetric behavior. This could be induced by the normal state background or the correlation effect[15,16,45,46] (see Supplementary Materials). In order to have a valid comparison with the theoretical fitting curves, we normalize the spectrum measured at 0.4 K with the one measured at 40 K, and the asymmetry becomes weaker. Since there is an obvious two-gap feature on the spectrum, we use the two-gap Dynes model[44] with different gap functions to fit the normalized spectrum, and the results are shown in Figs. 3c-3e. We first try the fitting with two components, each has an anisotropic *s*-wave gap. The fitting curve together with the normalized data are shown in Fig. 3c, and one can see that the data can be nicely fitted with $\Delta_{1S} = 18.8 \times (0.1\cos 4\theta + 0.9)$ meV and $\Delta_{2S} = 6 \times (0.2\cos 4\theta + 0.8)$ meV with a weight of 80% for the larger gap. Then we use an anisotropic *s*-wave gap for the larger one, and a *d*-wave gap for the smaller one; the used gap functions are $\Delta_{1s} = 18.8 \times (0.1\cos 4\theta + 0.9)$ meV and $\Delta_{2d} = 6\cos 2\theta$ meV. Now the weight for the larger gap is still about 80%. The results are shown in Fig. 3d. We must say the fitting has the same quality as the one by using two anisotropic *s*-wave gaps. In above two fittings, we use an anisotropic gap for the larger gap. The nice fittings yield a small anisotropy for the larger gap, at most about 20%. However, if we use a model of *d*-wave for the larger gap, it is impossible to get any reasonable fit by tuning all parameters. The best fit by using two *d*-wave gaps is shown in Fig. 3e, here we use the two gaps $\Delta_{1d} = 20\ \cos 2\theta$ meV and $\Delta_{2d} = 6\ \cos 2\theta$ meV. The main difficulty is that if a *d*-wave gap is taken for the larger gap value around 20 meV, it has no way to fit the global shape of the spectrum, and the kink corresponding to the small gap is completely smeared out. Based on the symmetry argument, it should not be possible to have the combination of an s-wave and a d-wave. In this sense, the only possibility for the former two case is that with two s-wave gaps, namely the s+-. The parameters for all fittings mentioned above are given in Extended Data Table I. Through our fitting, it is found that the broadening factor $\Gamma$ is quite large for the smaller gap, that could be the reason why we have a significant zero bias conductance. As a control experiment, similar conclusions for the superconducting gap structure and magnitude are obtained from the fitting results to the spectra measured in another area on the same sample through another tip-excavation process, see data in Extended Data Fig. 2 and Extended Data Fig. 3, as well as Extended Data Table II. The consistent gap values and conclusions obtained from two sets of independent measurements in different regions imply that the message revealed by our data are intrinsic.

## Discussion and comparison with theory

As shown above, in the superconducting $La_2PrNi_2O_7$ films, we have observed that the dominant gap (~ 19 meV) should have an *s*-wave symmetry with a small anisotropy. The smaller gap (~ 6-8 meV) should also have anisotropy and it is not discernable whether it is an anisotropic *s*-wave gap (probably with accidental nodes) or a *d*-wave gap. We also argued that by symmetry, it is not possible to have a combination of *s*-wave and *d*-wave, thus the most possible case would be two anisotropic *s*-wave gaps. Actually, the symmetry for the dominant gap is the most important factor to decide which pairing model is really functioning in the system. We start to discuss this issue by taking a simplified three-band model. For the $NiO_2$-bilayer system, the low-energy electronic structure is typically described by the low-energy effective model consisting of two $3d_{x^2-y^2}$ and $3d_{z^2}$ -orbitals located on two sites in a bilayer[17], as shown in Fig. 4a. Introducing bonding-antibonding combinations of the orbitals the normal state Hamiltonian can be diagonalized and according to the DFT results[17] there are three $\alpha$, $\beta$, and $\gamma$ bands, which cross the Fermi level. Within bonding-antibonding classification $\alpha$-, and $\gamma$-bands can be characterized as bonding bands of mostly $3d_{x^2-y^2}$ and $3d_{z^2}$ – orbital character at the Fermi level, respectively, while β-band is an antibonding one with $3d_{x^2-y^2}$ -mostly orbital-character at the Fermi surface, these three bands are schematically shown in Figs. 4b and 4c. Observe that α and β have a sizable bonding-antibonding splitting, which is a result of the admixture of the $3d_{z^2}$ – orbital to these bands. In addition, the antibonding $3d_{z^2}$ – mostly orbital band appears to be well above the Fermi level[47]. At the same time, the exact position of the bonding γ-band with respect to the Fermi level and its classification is not fully settled both experimentally[32,47] and theoretically[48]. It is interesting to note that theoretically the superconducting gap magnitudes on the bonding α and γ bands are very similar in magnitudes and of the same sign in $s^{\pm}$ scenario[34].

The superconducting gaps can be characterized by the intralayer gaps $\Delta_{||}(\boldsymbol{k})$ driven by the intralayer magnetic fluctuations and the interlayer gaps, $\Delta_{\perp}(\boldsymbol{k})$, driven by the interlayer magnetic fluctuations, supported by the strong hybridization of the $3d_{z^2}$ – orbitals between the layers. Overall, the superconducting state can be described by the following generic 4 × 4 Hamiltonian matrix in the Nambu-Gor'kov basis:

$$\widehat{H}_{SC} = \Phi^{\dagger} \begin{pmatrix} \varepsilon_{||}(\boldsymbol{k}) & t_{\perp}(\boldsymbol{k}) & \Delta_{||}(\boldsymbol{k}) & \Delta_{\perp}(\boldsymbol{k}) \\ t_{\perp}(\boldsymbol{k}) & \varepsilon_{||}(\boldsymbol{k}) & \Delta_{\perp}(\boldsymbol{k}) & \Delta_{||}(\boldsymbol{k}) \\ \Delta_{||}(\boldsymbol{k}) & \Delta_{\perp}(\boldsymbol{k}) & -\varepsilon_{||}(\boldsymbol{k}) & -t_{\perp}(\boldsymbol{k}) \\ \Delta_{\perp}(\boldsymbol{k}) & \Delta_{||}(\boldsymbol{k}) & -t_{\perp}(\boldsymbol{k}) & -\varepsilon_{||}(\boldsymbol{k}) \end{pmatrix} \Phi, \quad (1)$$

where $\Phi = \left(c_{1,k\uparrow}, c_{2,k\uparrow}, c_{1-k\downarrow}^{\dagger}, c_{2-k\downarrow}^{\dagger}\right)$ are the second quantization operators of layers 1 and 2, $\varepsilon_{||}(\boldsymbol{k})$ is the in-plane dispersion and $t_{\perp}(\boldsymbol{k})$ is the interlayer hybridization. In addition, each element of the matrix has $3d_{x^2-y^2}$ and $3d_{z^2}$ – orbital index as well. Most importantly, Eq. (1) can be diagonalized by transforming the basis into bonding-antibonding one with respect to the layer index $c_{b,k\uparrow} = \frac{1}{\sqrt{2}}\left(c_{1,k\uparrow} + c_{2,k\uparrow}\right)$ and $c_{a,k\uparrow} = \frac{1}{\sqrt{2}}\left(c_{1,k\uparrow} - c_{2,k\uparrow}\right)$ assuming inversion symmetry between the layers and the eigenenergies decompose into $E_b(\boldsymbol{k}) =$

$\pm\sqrt{\left(\varepsilon_b(\boldsymbol{k})\right)^2+\left(\Delta_b(\boldsymbol{k})\right)^2}$ and $E_a(\boldsymbol{k})=\pm\sqrt{\left(\varepsilon_a(\boldsymbol{k})\right)^2+\left(\Delta_a(\boldsymbol{k})\right)^2}$ where $\varepsilon_a(\boldsymbol{k})$ and $\varepsilon_b(\boldsymbol{k})$ refer to the bonding ($\alpha$ and $\gamma$) bands and antibonding $\beta$ band, which give rise to the $\alpha$, $\beta$, and $\gamma$ Fermi surface sheets in the normal state. Most importantly, the gaps on the $\alpha$ and $\gamma$ bands can be approximated by $\Delta_b(\boldsymbol{k})=\Delta_{||}(\boldsymbol{k})+\Delta_{\perp}(\boldsymbol{k})$ and on the $\beta$ band $\Delta_a(\boldsymbol{k})=\Delta_{||}(\boldsymbol{k})-\Delta_{\perp}(\boldsymbol{k})$. For further analysis we recall that the most common candidate solutions for the superconducting gap in the bilayer nickelates, as discussed in the literature, are either sign-changing bonding-antibonding $s^{\pm}$-wave symmetry or *d*-wave (either $d_{xy}$ or $d_{x^2-y^2}$) symmetry solutions[28], as illustrated in Figs. 4b and 4c. For the $s^{\pm}$-wave symmetry solution, it is important that the interlayer gap is expected to be larger in magnitude than the intralayer one, i.e. superconductivity is driven by the interlayer spin fluctuations stemming mostly from the interlayer coupling of the $3d_{z^2}$-orbtials. We further assume $\Delta_{\perp}$ to be momentum independent, while $\Delta_{||}^{s}(\boldsymbol{k})=\frac{\Delta_{s,||}^{0}}{2}(\cos k_x+\cos k_y)$. Depending on $\Delta_{s,||}^{0}$ magnitude the gaps become anisotropic and may even acquire accidental nodes on the $\beta$ band if intralayer and interlayer gaps become of the same order. At the same time the gap magnitudes on the $\alpha$ and $\gamma$ bands are found nearly constant and very similar to each other[34].

In the case of the *d*-wave symmetry, it is expected that these are derived from intralayer spin fluctuations, which mostly drive the Cooper-pairing. We proceed by assuming $d_{x^2-y^2}$-wave symmetry for concreteness yet $d_{xy}$-wave gives similar results. The intralayer gap is then $\Delta_{||}^{d}(\boldsymbol{k})=\frac{\Delta_{d,||}^{0}}{2}(\cos k_x-\cos k_y)$ and by symmetry arguments the interlayer gap should also acquire the *d*-wave form factor, i.e. $\Delta_{\perp}^{d}(\boldsymbol{k})=\frac{\Delta_{d,\perp}^{0}}{2}(\cos k_x-\cos k_y)$ and $\Delta_{d,||}^{0}>\Delta_{d,\perp}^{0}$. Most importantly the non-zero $\Delta_{d,\perp}^{0}$ value yields different magnitude of gaps on the $\alpha$, $\beta$, and $\gamma$ Fermi surface sheets but do not introduce any additional anisotropy. The latter would in principle appear if one introduces higher in-plane harmonics of the *d*-wave symmetry.

In the next step we assume that the gaps on the $\alpha$ and $\gamma$-bands are either of the same magnitudes[19,34] or the $\gamma$-band is well below the Fermi level and does not contribute to the superfluid density although it participates in the pairing, while the smaller gap resides on the $\beta$ band. Keeping in mind that the superconducting gaps can be treated independently on different bands, we thus proceed with fitting the experimental data using the Dynes formula with two gaps assuming three different scenarios: (i) two anisotropic *s*-wave gaps, (ii) one anisotropic *s*-wave and one *d*-wave, and (iii) two *d*-wave superconducting gaps of different magnitudes, as proposed in some models[49,50]. Our fitting to the experimental data, as discussed above, clearly indicates that the third case, namely the global *d*-wave gap cannot fit the experimental data at all. As our experimental data support the dominance of anisotropic *s*-wave gaps in the present system, we can estimate the sizes of the interlayer and intralayer gaps. Assuming that $\Delta_1$ and $\Delta_2$ have opposite signs, we obtain the magnitude of the interlayer gap to be $\Delta_\perp \approx 13$ meV, while the intralayer gap is $\Delta_{||} \approx 6$ meV. This can be taken as an estimate of the strength of the interlayer and intralayer spin fluctuations, respectively. We note, however, that the overall conclusion on the sign-changing bonding-antibonding $s^{\pm}$-wave symmetry needs to be verified by momentum resolved measurements in future experiments. In addition, the superconducting transition $T_c^{onset}$ is about 41.5 K in the present sample, which is still below the highest value reported for the system under pressure. Thus, it may be subject to some modifications in the future for samples with optimized superconductivity. Our present results give strong evidence that the dominant superconducting gap is induced by the interlayer pairing due to strong interlayer hybridization of the $3d_{z^2}$ – orbital yielding strong bilayer splitting of the $\alpha$ and $\beta$ Fermi surface pocket with or without the involvement of the $\gamma$ Fermi pocket. In the former case the superconducting gap magnitude on the $\gamma$-pocket should be the same as on the $\alpha$-pocket.

In conventional superconductors, the bosonic mode exhibits as a weak modification to the tunneling spectrum due to the interactions between the Bogoliubov quasiparticles and phonons, leading to some tiny kinky structures outside the coherence peaks[41,42]. However, the bosonic mode shows up as pronounced peaks in the tunneling spectrum or the neutron resonance peaks in some unconventional superconductors[51,52], because it reflects the interaction between the Bogoliubov quasiparticles and the resonance mode of spin-1 excitations in superconductors with reversal superconducting gaps. Observe that for the sign-changing bonding-antibonding $s^{\pm}$-wave symmetry the spin resonance would also form in the odd channel of the spin susceptivity, which would be a hallmark of this superconducting state[28]. From our spectrum measured at 0.4 K, the characteristic energy $\Delta+\Omega$ can be obtained at about 49±2 meV, using the larger gap value $\Delta_1$ of about 19 meV, the mode energy $\Omega$ is about 30±2 meV. Theoretically, it was predicted that the $\Omega \leq 2\Delta_1$ for unconventional superconductivity, which is just satisfied by our data. Usually there is a ratio between the mode energy and $T_c$, i.e. $\Omega/k_BT_c$ which locates in the range of 4-8[43,53]. In our present case, it gives a value of about $\Omega/k_BT_c = 8.4$, which suggests a strong coupling of the Cooper pairing induced by magnetic fluctuations. We think the peaks at ±49 meV are unlikely derived from the coherence peaks of the $\gamma$ band if it would appear as an incipient band below the Fermi level, since these two peaks show quite strong intensities. This feature is quite similar to the situation in iron based superconductors in which these peaks outside the superconducting coherence peaks were found to be associated well with the neutron resonance peaks, thus they correspond to the bosonic mode[43].

Our combined observations of *s*-wave symmetry for the larger gap and the bosonic mode point to the fact that the superconductivity is derived from the magnetic spin fluctuations with a sign reversal gap, namely the $s^{\pm}$ pairing. We note, however, that the overall conclusion on the sign-changing bonding-antibonding $s^{\pm}$-wave symmetry needs to be verified by momentum resolved measurements in future experiments. In addition, the superconducting transition $T_c^{onset}$ is about 41.5 K in the present sample, which is still below the highest value reported for the system under pressure. Thus, it may be subject to some modifications in the future for samples with optimized superconductivity. Our present results give strong evidence that the dominant superconducting gap is induced by the interlayer pairing even without the direct involvement of the $\gamma$ Fermi pocket at the Fermi level.

## Methods

### Film growth and characterization

The 2-unit-cell $La_2PrNi_2O_7$ thin films were epitaxially grown on single-crystal $SrLaAlO_4$ (001) substrates by pulsed laser deposition (PLD) using a KrF excimer laser ($\lambda$ = 248 nm). A polycrystalline $La_2PrNi_2O_7$ target, prepared via the sol-gel method, was ablated at a laser fluence of 1.2-1.4 J/cm$^2$ with a repetition rate of 2 Hz. The deposition was performed at 780°C under an oxygen partial pressure of 20 Pa. Following deposition, the sample temperature was cooled to 550°C at a controlled rate of 10°C/min in the same oxygen atmosphere. The samples were maintained at 550°C and then post-annealed with ~7 wt% ozone for 30 minutes. The ozone was delivered through a nozzle positioned 1.8 cm from the sample surface, with the partial pressure and flow rate precisely maintained at 20 Pa and 10 sccm, respectively. After ozone treatment, the temperature was cooled down to 250°C at a rate of 15°C/min, after which the ozone supply was turned off while maintaining the chamber pressure for natural cooling. The electrical transport properties were characterized using the standard four-probe method in a Physical Property Measurement System (PPMS, Quantum Design). Structural characterization was performed using a Bruker D8 Advanced X-ray diffractometer, and surface morphology was investigated by atomic force microscopy (FM-Nanoview6800).

**STM/STS measurements**

The STM/STS measurements were carried out in a scanning tunneling microscope (USM-1300, Unisoku Co., Ltd.) with ultra-high vacuum, low temperature, and high magnetic field. The thin films were transferred to the STM head across the high-vacuum chamber as soon as possible. But because our PLD machine and STM are not vacuum connected, the films were exposed to air before they were put into the STM chamber. The electrochemically etched tungsten tips were used for the STM/STS measurements. A typical lock-in technique was used for the tunneling spectrum measurements with an ac modulation of 1% of $V_{set}$ and 931.773 Hz. For each spectrum shown in the work, in order to lower the noise, we measure six times at the same position and take the average of them as the final data shown here. Unless otherwise specified, the setpoint conditions for the tunneling spectrum measurements are: $V_{\mathrm{set}}$ = 50 mV and $I_{\mathrm{set}}$ = 100 pA.

**Dynes model fitting**

Based on the Dynes model, we describe the differential conductance as,

$$G(V) \propto \frac{d}{dV}\left(\int_{-\infty}^{+\infty} d\varepsilon \int_{0}^{2\pi} d\theta [f(\varepsilon) - f(\varepsilon + eV)] \cdot Re\left(\frac{\varepsilon + eV + i\Gamma}{\sqrt{(\varepsilon + eV + i\Gamma)^2 - \Delta^2(\theta)}}\right)\right) \qquad (2)$$

where *f(ε)* is the fermi distribution function, Δ denotes the superconducting gap value, and Γ represents the scattering factor assumed to be isotropic. The spectra are fitted by the Dynes model with two isolated superconducting gaps. In this situation, the total differential conductance can be expressed as $G = xG_1 + (1-x)G_2$, where $G_1$ or $G_2$ is the differential conductance contributed by the gap $\Delta_1$ or $\Delta_2$, and $x$ or $1-x$ is the spectral weight contributed by the gap $\Delta_1$ or $\Delta_2$. All parameters used for fitting are summarized in Extended Data Table I and Extended Data Table II for two different regions with distinct tip-excavation processes.

## Data availability

The data that support the findings of this study are available from the corresponding authors upon reasonable request.

## Acknowledgements

We appreciate the useful discussions with Harold Hwang in Stanford university and Daoxin Yao in Sun Yat-Sen university. This work is supported by the National Key R&D Program of China (Grant No. 2022YFA1403201 and No. 2024YFA1408104), the Natural Science Foundation of China (Grant No. 12494591, No. 11927809 and No. 12434004), and the Natural Science Foundation of Jiangsu Province (Grant No. BK20233001).

## Author contributions

The $La_2PrNi_2O_7$ thin films were grown and characterized by M.O., Q.L., Y.W., and H.-H.W. The STM measurements and data analysis were done by S.F., Z.S., J.X., H.Y., and H.H.W. Theoretical calculations were carried out by M.S. and I.E. The manuscript was written by S.F., Q.L., H.Y., I.E., and H.-H.W., which is supplemented by others. H.-H.W. has coordinated the whole work.

## Competing interests

The authors declare no competing interests.

## References

1. Li, D. et al. Superconductivity in an infinite-layer nickelate. *Nature* **572**, 624−627 (2019).
2. Zeng, S. et al. Phase diagram and superconducting dome of infinite-layer $Nd_{1-x}Sr_xNiO_2$ thin films. *Phys. Rev. Lett*. **125** 147003 (2020).

3. Gu, Q. & Wen, H.-H. Superconductivity in nickel-based 112 systems. *Innovation* **3**, 100202 (2022).

4. Sun, H. et al. Signatures of superconductivity near 80 K in a nickelate under high pressure. *Nature* **621**, 493-498 (2023).

5. Zhang, Y. et al. High-temperature superconductivity with zero resistance and strange-metal behaviour in $La_3Ni_2O_{7-\delta}$. *Nat. Phys.* **20**, 1269-1273 (2024).

6. Wang, N. et al. Bulk high-temperature superconductivity in pressurized tetragonal $La_2PrNi_2O_7$. *Nature* **634**, 579-584 (2024).

7. Li, F. et al. Ambient pressure growth of bilayer nickelate single crystals with superconductivity over 90 K under high pressure. Preprint at https://arxiv.org/abs/2501.14584 (2025).

8. Zhu, Y. et al. Superconductivity in pressurized trilayer $La_4Ni_3O_{10-\delta}$ single crystals. *Nature* **631**, 531-536 (2024).

9. Li, Q. et al. Signature of superconductivity in pressurized $La_4Ni_3O_{10}$. *Chin. Phys. Lett.* **41**, 017401 (2024).

10. Sakakibara, H. et al. Theoretical analysis on the possibility of superconductivity in the trilayer Ruddlesden-Popper nickelate $La_4Ni_3O_{10}$ under pressure and its experimental examination: Comparison with $La_3Ni_2O_7$. *Phys. Rev. B* **109**, 144511 (2024).

11. Shi, M. et al. Superconductivity of the hybrid Ruddlesden-Popper $La_5Ni_3O_{11}$ single crystals under high pressure. Preprint at https://arxiv.org/abs/2502.01018 (2025).

12. Ko, E. K. et al. Signatures of ambient pressure superconductivity in thin film $La_3Ni_2O_7$. *Nature* **638**, 935-940 (2025).

13. Zhou. G. et al. Ambient-pressure superconductivity onset above 40 K in $(La,Pr)_3Ni_2O_7$ films. *Nature* **640**, 641-646 (2025).

14. Bao, B. et al. Superconductivity and phase diagram in Sr-doped $La_3Ni_2O_7$ thin films.

Preprint at https://arxiv.org/abs/2505.12603 (2025).

15. Yang, J. et al. Orbital-dependent electron correlation in double-layer nickelate $La_3Ni_2O_7$. *Nat. Commun.* **15**, 4373 (2024).

16. Abadi, S. et al. Electronic structure of the alternating monolayer-trilayer phase of $La_3Ni_2O_7$. *Phys. Rev. Lett.* **134**, 126001 (2025).

17. Luo, Z., Hu, X., Wang, M., Wú, W. & Yao, D. X. Bilayer two-orbital model of $La_3Ni_2O_7$ under pressure. *Phys. Rev. Lett.* **131**, 126001 (2023).

18. Zhang, Y., Lin, L. F., Moreo, A. & Dagotto, E. Electronic structure, dimer physics, orbital-selective behavior, and magnetic tendencies in the bilayer nickelate superconductor $La_3Ni_2O_7$ under pressure. *Phys. Rev. B* **108**, L180510 (2023).

19. Yang, Q. G., Wang, D. & Wang, Q. H. Possible $s_{\pm}$-wave superconductivity in $La_3Ni_2O_7$. *Phys. Rev. B* **108**, L140505 (2023).

20. Shen, Y., Qin, M. & Zhang, G. M. Effective bi-layer model Hamiltonian and density-matrix renormalization group study for the high-$T_c$ superconductivity in $La_3Ni_2O_7$ under high pressure. *Chin. Phys. Lett.* **40**, 127401 (2023).

21. Lechermann, F., Gondolf, J., Bötzel, S. & Eremin, I. M. Electronic correlations and superconducting instability in $La_3Ni_2O_7$ under high pressure. *Phys. Rev. B* **108**, L201121 (2023).

22. Qin, Q. & Yang, Y. F. High-$T_c$ superconductivity by mobilizing local spin singlets and possible route to higher $T_c$ in pressurized $La_3Ni_2O_7$. *Phys. Rev. B* **108**, L140504 (2023).

23. Christiansson, V., Petocchi, F. & Werner, P. Correlated electronic structure of $La_3Ni_2O_7$ under pressure. *Phys. Rev. Lett.* **131**, 206501 (2023).

24. Yang, Y. F., Zhang, G. M. & Zhang, F. C. Interlayer valence bonds and two-component theory for high-$T_c$ superconductivity of $La_3Ni_2O_7$ under pressure. *Phys. Rev. B* **108**, L201108 (2023).

25. Liu, Y. B., Mei, J. W., Ye, F., Chen, W. Q. & Yang, F. $s^{\pm}$-wave pairing and the destructive role of apical-oxygen deficiencies in $La_3Ni_2O_7$ under pressure. *Phys. Rev. Lett.* **131**, 236002 (2023).

26. Luo, Z., Lv, B., Wang, M., Wú, W. & Yao, D.-X. High-$T_C$ superconductivity in $La_3Ni_2O_7$ based on the bilayer two-orbital t-J model. *npj Quan. Mater.* **9**, 61 (2024).

27. Ouyang, Z., Gao, M. & Lu, Z. Y. Absence of electron-phonon coupling superconductivity in the bilayer phase of $La_3Ni_2O_7$ under pressure. *npj Quan. Mater.* **9**, 80 (2024).

28. Bötzel, S., Lechermann, F., Gondolf, J. & Eremin, I. M. Theory of magnetic excitations in the multilayer nickelate superconductor. *Phys. Rev. B* **109**, L180502 (2024).

29. Xi, W., Yu, S.-L. & Li, J.-X., Transition from $s_{\pm}$-wave to $d_{x^2-y^2}$-wave superconductivity driven by interlayer interaction in the bilayer two-orbital model of $La_3Ni_2O_7$. *Phys. Rev. B* **111**, 104505 (2025).

30. Gu, Y., Le, C., Yang, Z., Wu, X. & Hu. J. Effective model and pairing tendency in the bilayer Ni-based superconductor. *Phys. Rev. B* **111**, 174506 (2025).

31. Jiang, K.-Y., Cao, Y.-H., Yang, Q.-G., Lu, H.-Y. & Wang, Q.-H. Theory of pressure dependence of superconductivity in bilayer nickelate $La_3Ni_2O_7$. *Phys. Rev. Lett.* **134**, 076001 (2025).

32. Wang, B. Y. et al., Electronic structure of compressively strained thin film $La_2PrNi_2O_7$. Preprint at https://arxiv.org/abs/2504.16372 (2025).

33. Li, P. et al. Angle-resolved photoemission spectroscopy of superconducting $(La,Pr)_3Ni_2O_7/SrLaAlO_4$ heterostructures. *Natl. Sci. Rev.* nwaf205 (2025). DOI: 10.1093/nsr/nwaf205.

34. Zhang, Y., Lin, L.-F., Moreo, A., Maier, T. A. & Dagotto, E. Structural phase transition, $s_{\pm}$-wave pairing, and magnetic stripe order in bilayered superconductor $La_3Ni_2O_7$ under pressure. *Nature Communications* **15**, 2470 (2024).

35. Fan, Z. et al. Superconductivity in nickelate and cuprate superconductors with strong bilayer coupling. *Phys. Rev. B* **110**, 024514 (2024).

36. Jiang, K., Wang, Z. & Zhang, F.-C. High-temperature superconductivity in $La_3Ni_2O_7$. *Chin. Phys. Lett.* **41**, 017402 (2024).

37. Xia, C., Liu, H., Zhou, S. & Chen, H. Sensitive dependence of pairing symmetry on Ni-$e_g$ crystal field splitting in the nickelate superconductor $La_3Ni_2O_7$. *Nat. Commun.* **16**, 1054 (2025).

38. Liu, C. et al. Andreev reflection in superconducting state of pressurized $La_3Ni_2O_7$. *Sci. China-Phys. Mech. Astron.* **68**, 247412 (2025).

39. Shen, J. et al. Anomalous energy gap in superconducting $La_{2.85}Pr_{0.15}Ni_2O_7/SrLaAlO_4$ heterostructures. Preprint at https://arxiv.org/abs/2502.17831 (2025).

40. Gu, G. et al. Single particle tunneling spectrum of superconducting $Nd_{1-x}Sr_xNiO_2$ thin films, *Nat. Commun.* **11**, 6027 (2020).

41. Rowell, J. M., Anderson, P. W. & Thomas, D. E. Image of the phonon spectrum in the tunneling characteristic between superconductors. *Phys. Rev. Lett.* **10**, 334-336 (1963).

42. Schrieffer, J. R., Scalapino, D. J. & Wilkins, J. W. Effective tunneling density of states in superconductors. *Phys. Rev. Lett.* **10**, 336-339 (1963).

43. Wang, Z. et al. Close relationship between superconductivity and the bosonic mode in $Ba_{0.6}K_{0.4}Fe_2As_2$ and $Na(Fe_{0.975}Co_{0.025})As$. *Nat. Phys.* **9**, 42-48 (2013)

44. Dynes, R. C., Garno, J. P., Hertel, G. B. & Orlando, T. P. Tunneling study of superconductivity near the metal-insulator transition. *Phys. Rev. Lett.* **53**, 2437 (1984).

45. Liu, Z. et al. Electronic correlations and partial gap in the bilayer nickelate $La_3Ni_2O_7$. *Nat. Commun.* **15**, 7570 (2020).

46. Fan, S. et al. Tunneling spectra with gaplike features observed in nickelate $La_3Ni_2O_7$ at ambient pressure. *Phys. Rev. B* **110**, 134520 (2024).

47. Li, P. *et al.* Photoemission evidence for multi-orbital hole-doping in superconducting $La_{2.85}Pr_{0.15}Ni_2O_7/SrLaAlO_4$ interfaces. Preprint at https://arxiv.org/abs/2501.09255 (2025).
48. Lechermann, F., Bötzel, S. & Eremin, I. M. Low-energy perspective of interacting electrons in the normal state of superconducting bilayer nickelate. Preprint at https://arxiv.org/abs/2503.12412 (2025).
49. Singh, D. K., Goyal, G. & Bang, Y. Possible pairing states in the superconducting bilayer nickelate. *New J. Phys.* **27**, 053503 (2025).
50. Wang, Z., Zhang, G.-M., Yang, Y.-f. & Zhang, F.-C. Distinct pairing symmetries of superconductivity in infinite-layer nickelates. *Phys. Rev. B* **102**, 220501 (2020).
51. Scalapino, D. J. A common thread: The pairing interaction for unconventional superconductors. *Rev. Mod. Phys.* **84**, 1383-1417 (2012).
52. Christianson, A. D. et al. Unconventional superconductivity in $Ba_{0.6}K_{0.4}Fe_2As_2$ from inelastic neutron scattering. *Nature* **456**, 930-932 (2008).
53. Wilson, S. D. et al. Resonance in the electron-doped high-transition-temperature superconductor $Pr_{0.88}LaCe_{0.12}CuO_{4-\delta}$. *Nature* **442**, 59-62 (2006).

## Figures

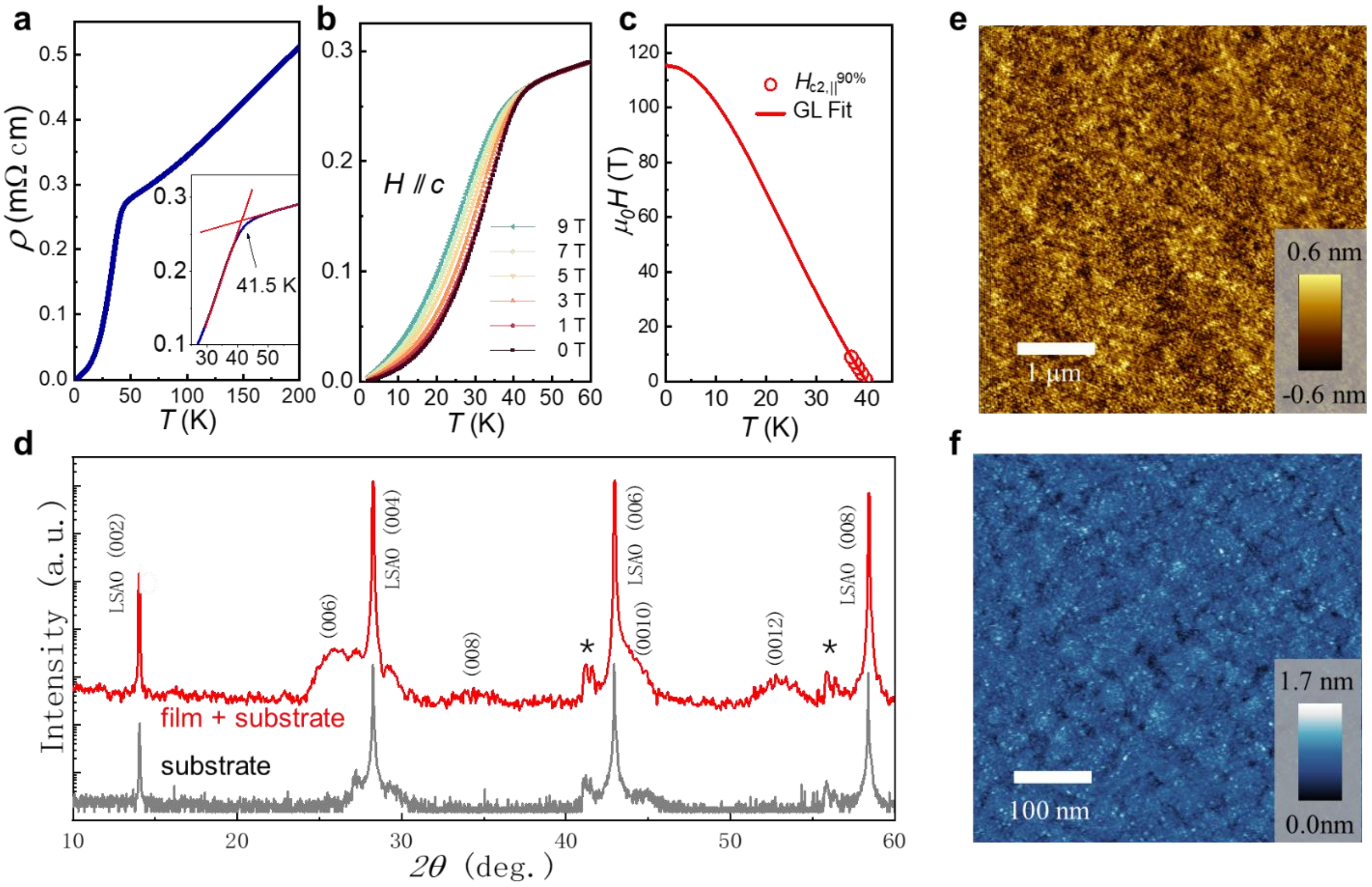

**Fig. 1 | Characterization of superconducting $La_2PrNi_2O_7$ thin films. a**, Temperature dependent resistivity of the $La_2PrNi_2O_7$ thin film. The inset shows the determination of $T_c^{onset}$, and the temperature is about 41.5 K. **b**, $\rho$-T curves of the $La_2PrNi_2O_7$ thin film under various magnetic fields applied along the *c*-axis. **c**, Temperature dependent upper critical fields extracted by the criteria of 90%$\rho_n(T)$. The experimental data are fitted by the formula $\mu_0 H_{c2}(T) = \mu_0 H_{c2}(0)\,[1 - (T/T_c)^2]/[1 + (T/T_c)^2]$, and the obtained $\mu_0 H_{c2}(0)$ =115 T. **d**, X-ray diffraction data of 2θ-ω scans of a 2UC $La_2PrNi_2O_7$ thin film grown on $SrLaAlO_4$ substrate. The data of the pure $SrLaAlO_4$ substrate are also provided for comparison. The tiny peaks marked with asterisks seem to be derived from the substrate. **e**, Atomic force microscopy image of $La_2PrNi_2O_7$ thin film. **f**, Topographic image recorded by STM in an area of 400 × 400 $nm^2$ (Setpoint conditions: $V_{set}$ = 8 V and $I_{set}$ = 20 pA).

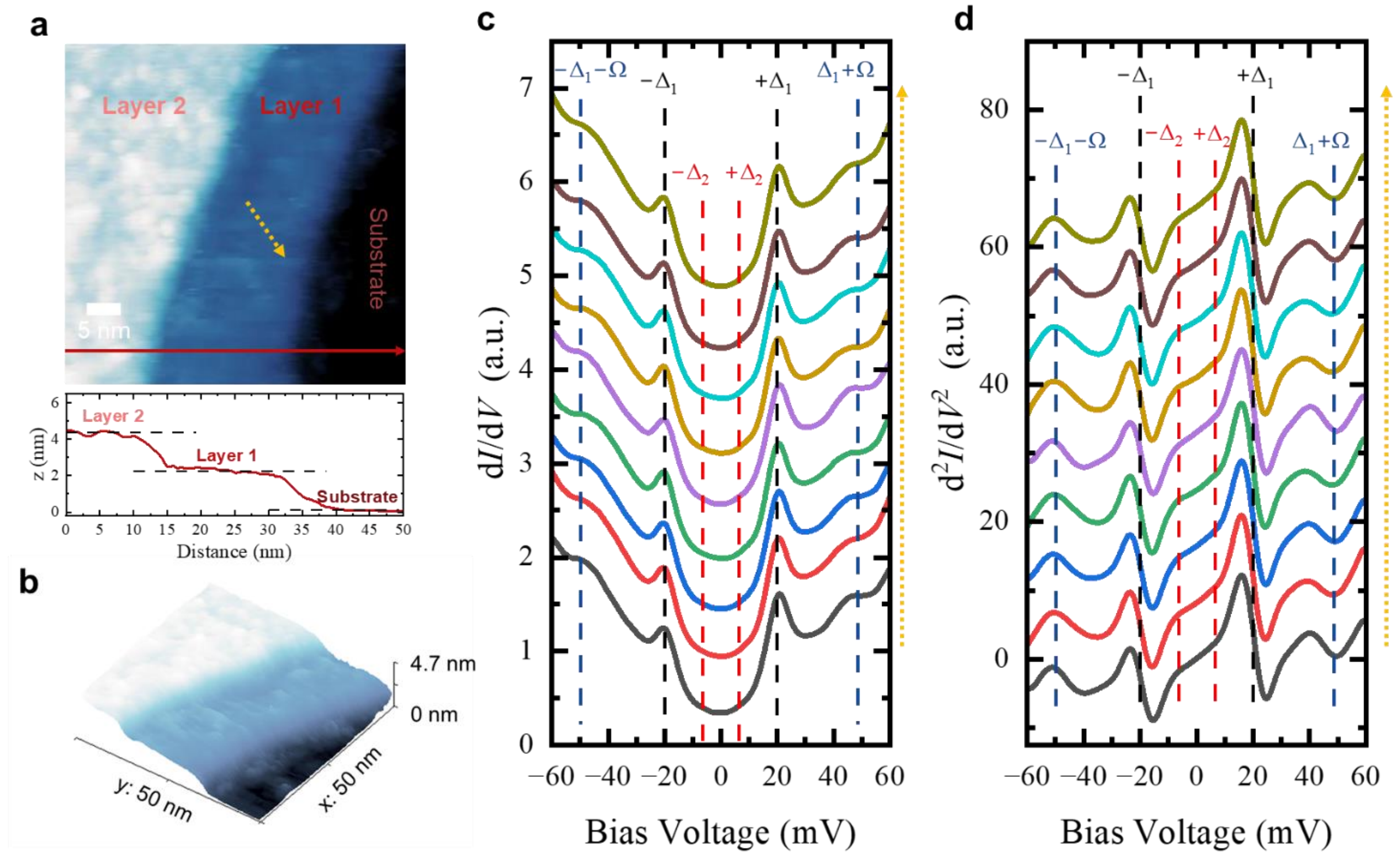

**Fig. 2 | Superconducting region obtained after the tip-excavation and the tunneling spectrum. a**, Topographic image in an area after applying a bias pulse of 4 V to the tip when it is stabilized in the setpoint conditions of $V_{set}$ = 8 V and $I_{set}$ = 10 pA. Some atomic layers of the film are removed by the operation forming some terraces. Below shows the height distribution measured along the red line in the topographic image. The two steps correspond to 1UC thickness differences of different layers. **b**, Three-dimensional plot of the height shown in **a**. It clearly illustrates the height variations and the terraces. **c**, A set of tunneling spectra measured along the dotted arrowed line in Layer 1. The spectra show the two-gap features which are illustrated by the vertical red and black dashed lines. **d**, The second derivative curves $d^2I/dV^2$ versus $V$ correspond to the spectra in **c**. The energy for the bosonic mode can be determined at the local maximum (minimum) of the $d^2I/dV^2$ curves at negative (positive bias). The energies determined corresponding to the bosonic mode characteristics $\pm(\Delta+\Omega)$ are about ±49 meV.

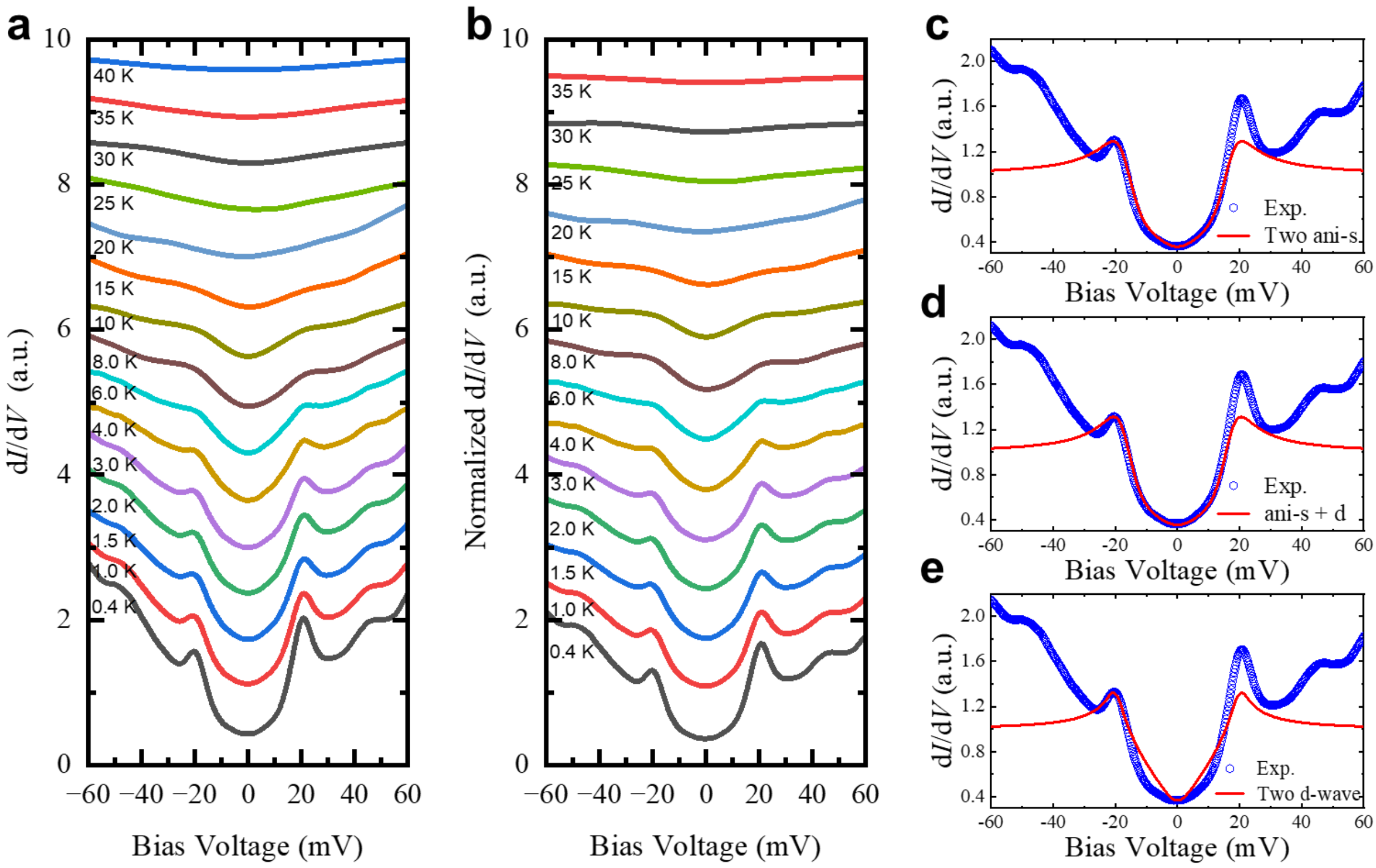

**Fig. 3 | Tunneling spectra measured at different temperatures and the fitting results. a**, Tunneling spectra measured at different temperatures. **b**, Tunneling spectra in **a** normalized by the spectrum measured at 40 K as the background. **c**-**e**, Theoretical fitting results by Dynes model with two gaps to the normalized spectrum at 0.4 K. The gap functions used in the fittings are: **c**, $\Delta_{1s}(\theta) = 18.8(0.9+0.1\cos4\theta)$ meV, and $\Delta_{2s}(\theta) = 6(0.8+0.2\cos4\theta)$ meV; **d**, $\Delta_{1s}(\theta) = 18.8(0.9+0.1\cos4\theta)$ meV, and $\Delta_{2d}(\theta) = 6\cos2\theta$ meV; **e**, $\Delta_{1d}(\theta) = 20\cos2\theta$ meV, and $\Delta_{2d}(\theta) = 6\cos2\theta$ meV. Obviously, the fitting results in **c** and **d** are in good agreement with the measured spectrum, while the fitting in **e** fails to catch up main features of the experimental data.

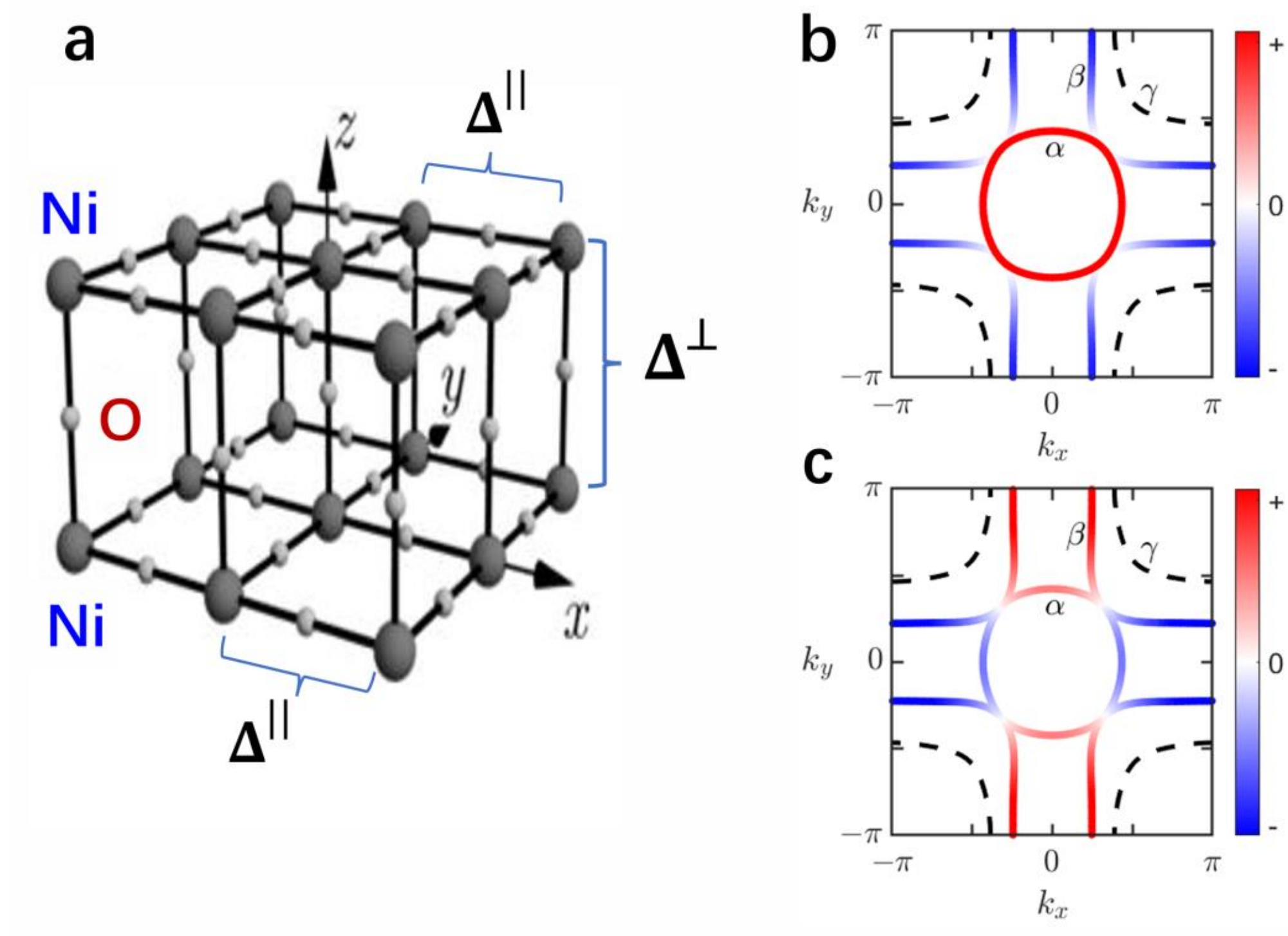

**Fig. 4 | Theoretical modeling and the illustration of the superconducting gaps. a,** Skeletons of the bilayer Ni-O blocks. Two NiO planes share the apical oxygen atoms**. b**,**c**, The Fermi surfaces and two dominant pairing models: **b** for $s^{\pm}$ and **c** for *d*-wave. According to our observation and related analysis, the $\gamma$ pocket is either having the same superconducting gap magnitude as on the $\alpha$-pocket or does not contribute to the superfluid density because it is well below the Fermi level, although it may still contribute to the Cooper pairing.

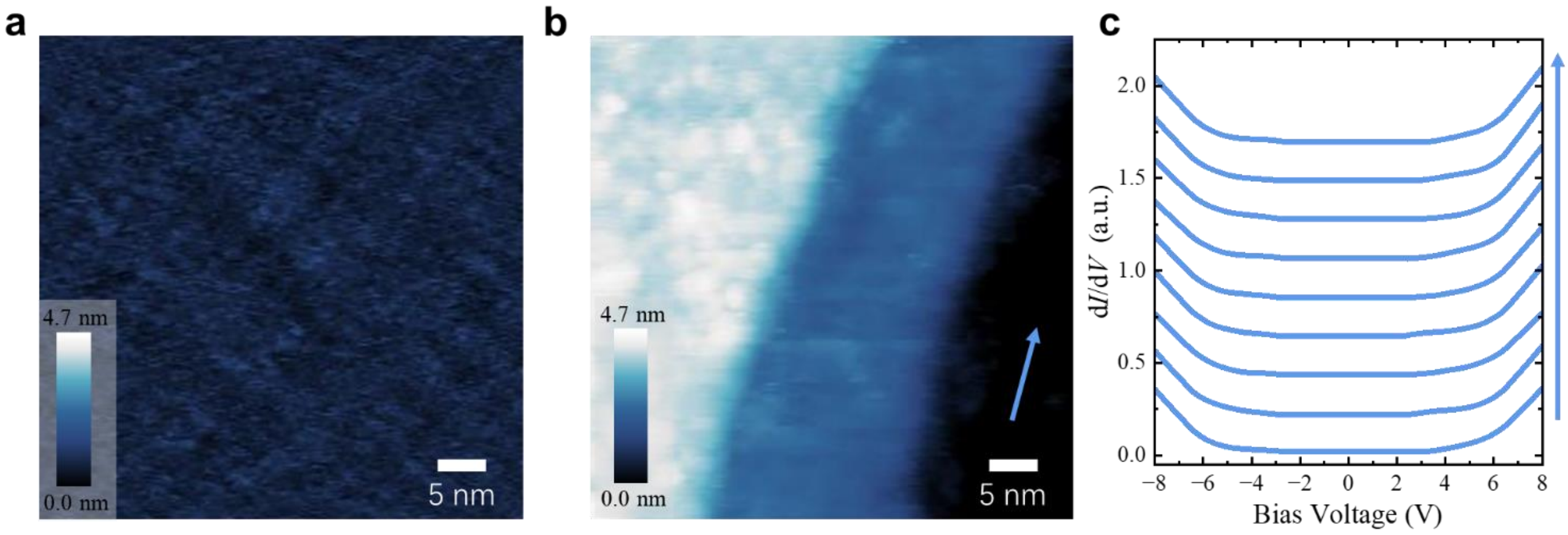

**Extended Data Fig. 1 | Topographic image before and after the tip-excavation treatment.** **a**, Topographic image in an area before the tip-excavation. **b**, Topographic image in the same area after the tip-excavation, two steps are exposed by this tip-excavation, each step corresponds just one unit cell of the film, the dark area reflects the substrate. **c**, A set of tunneling spectra measured along the blue arrowed line in the substrate layer. The spectra show a very large insulating gap. (Setpoint conditions for **a** and **b**: $V_{set}$ = 8 V and $I_{set}$ = 20 pA; for **c**: $V_{set}$ = 8 V and $I_{set}$ = 100 pA).

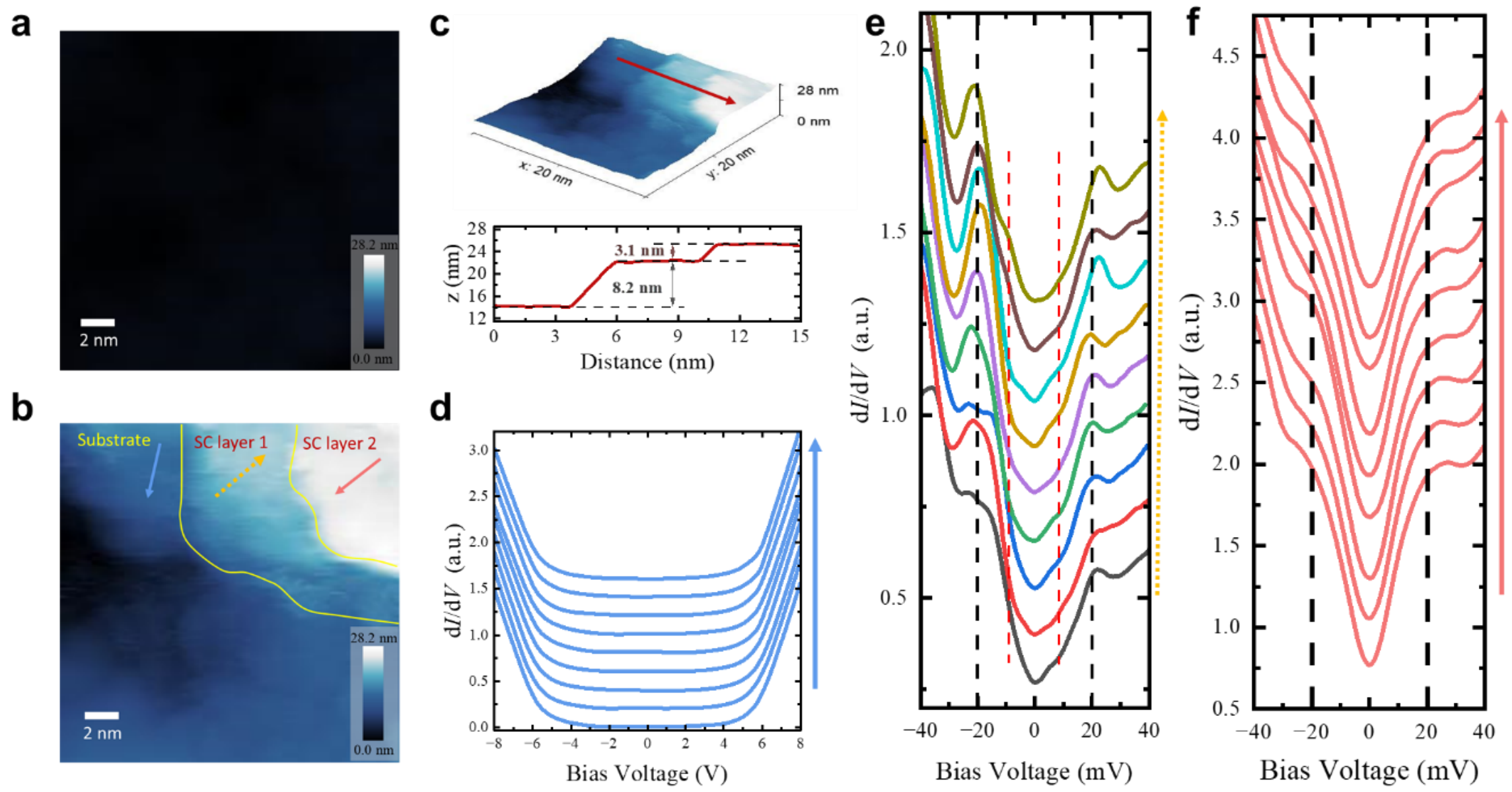

**Extended Data Fig. 2 | Morphology and spectra obtained in another region. a**,**b**, Topographic images in another area before and after the tip-excavation, respectively. Some atomic layers of the film and even some layers of the surface substrate are removed by the operation forming some terraces. There are two terrace layers (SC layer 1 and SC layer 2) with different heights, both showing superconducting features. The two yellow lines show the rough positions of the terrace edges. **c**, Three-dimensional plot of the height shown in **b**. It more clearly illustrates the height variations and the terraces. Height distribution measured along the red line in the three-dimensional plot. The two steps correspond to different film surfaces. **d**, A set of tunneling spectra measured along the blue arrowed line in the substrate. The spectra show a very large insulating gap. **e**, A set of tunneling spectra measured along the dotted arrowed line in SC layer 1. The spectra show the two-gap features which are illustrated by the vertical dashed lines. **f**, A set of tunneling spectra measured on the superconducting layer 2 along the arrowed line, the gapped feature at about 20 meV is still visible, but the coherence peaks are almost invisible. (Setpoint conditions for **a** and **b**: $V_{set}$ = 8 V and $I_{set}$ = 20 pA; for **d**: $V_{set}$ = 8 V and $I_{set}$ = 100 pA; )

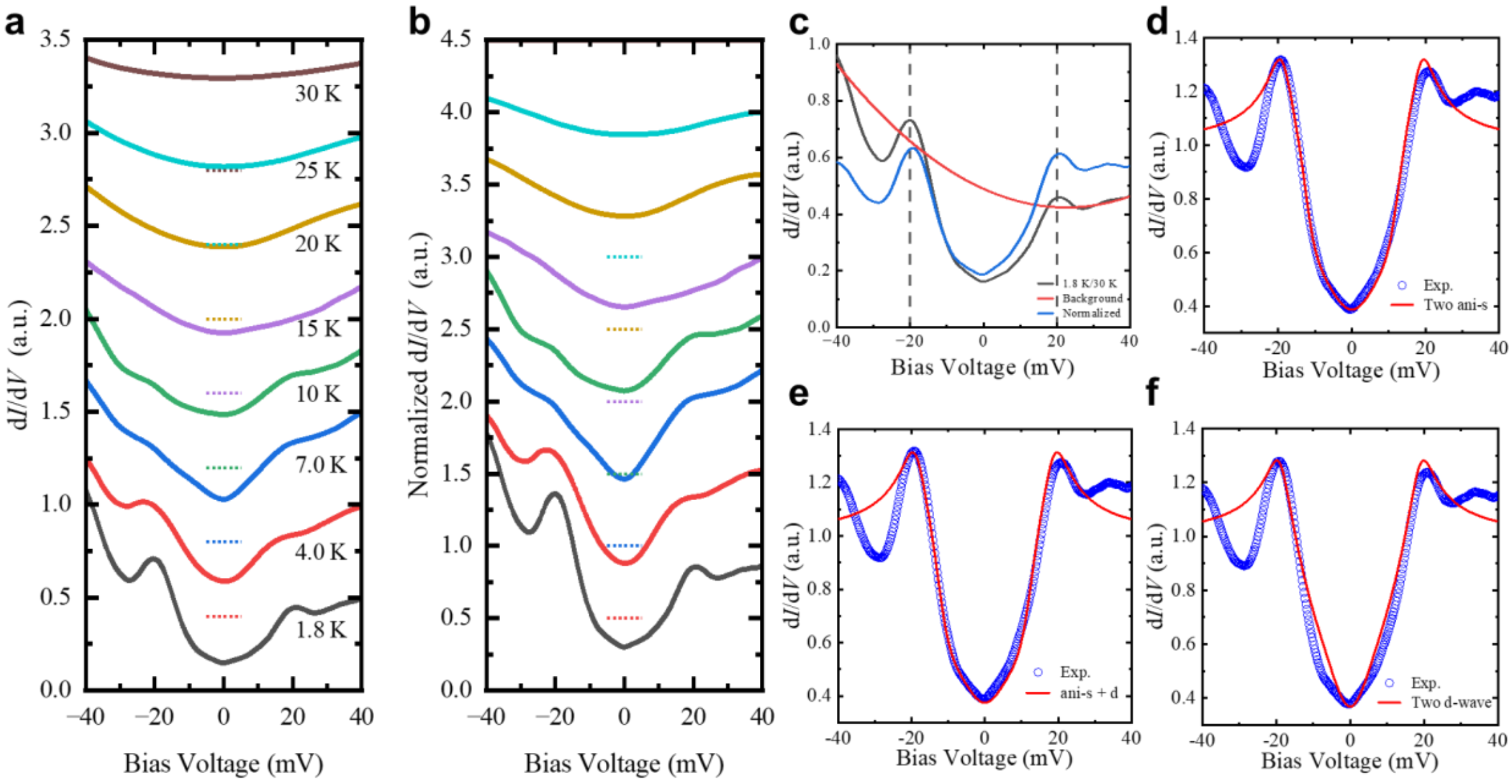

**Extended Data Fig. 3 | Tunneling spectra measured at different temperatures and the fitting results for another region. a**, Tunneling spectra measured at different temperatures as that shown in Extended Data Fig. 2. **b**, Tunneling spectra normalized by the spectrum measured at 30 K as the background. **c**, Spectra before and after the subtraction of a background. After the normalization of the spectrum measured at 1.8 K by that at 30 K, there is still an inclined background. We thus fit the spectrum beyond the range of ±35 mV with an inclined function $y = ax^2 + bx + c$, here x = bias voltage in unit of mV, y = d*I*/d*V*, the obtained values are $a = 1.3\times10^{-4}$, $b = -5.9\times10^{-3}$, $c = 0.49$. The obtained spectrum shows almost particle-hole symmetric behavior. **d-f**, Theoretical fitting results by Dynes model with two gaps to the normalized spectrum at 1.8 K. The gap functions used in the fittings are: **d**, two anisotropic *s*-wave gaps: $\Delta_{1s}(\theta)$ = 19(0.83+0.17cos4$\theta$) meV, and $\Delta_{2s}(\theta)$ = 6(0.8+0.2cos4$\theta$) meV; **e**, one anisotropic *s*-wave gap and one *d*-wave gap: $\Delta_{1s}(\theta)$ = 19(0.82+0.18cos4$\theta$) meV, and $\Delta_{2d}(\theta)$ = 5.5cos2$\theta$ meV; **f**, two *d*-wave gaps: $\Delta_{1d}(\theta)$ = 19cos2$\theta$ meV, and $\Delta_{2d}(\theta)$ = 6cos2$\theta$ meV. Obviously, the fitting results in **d** and **e** are in good agreement with the measured spectrum, while the fitting in **f** fails to catch up main features of the experimental data.

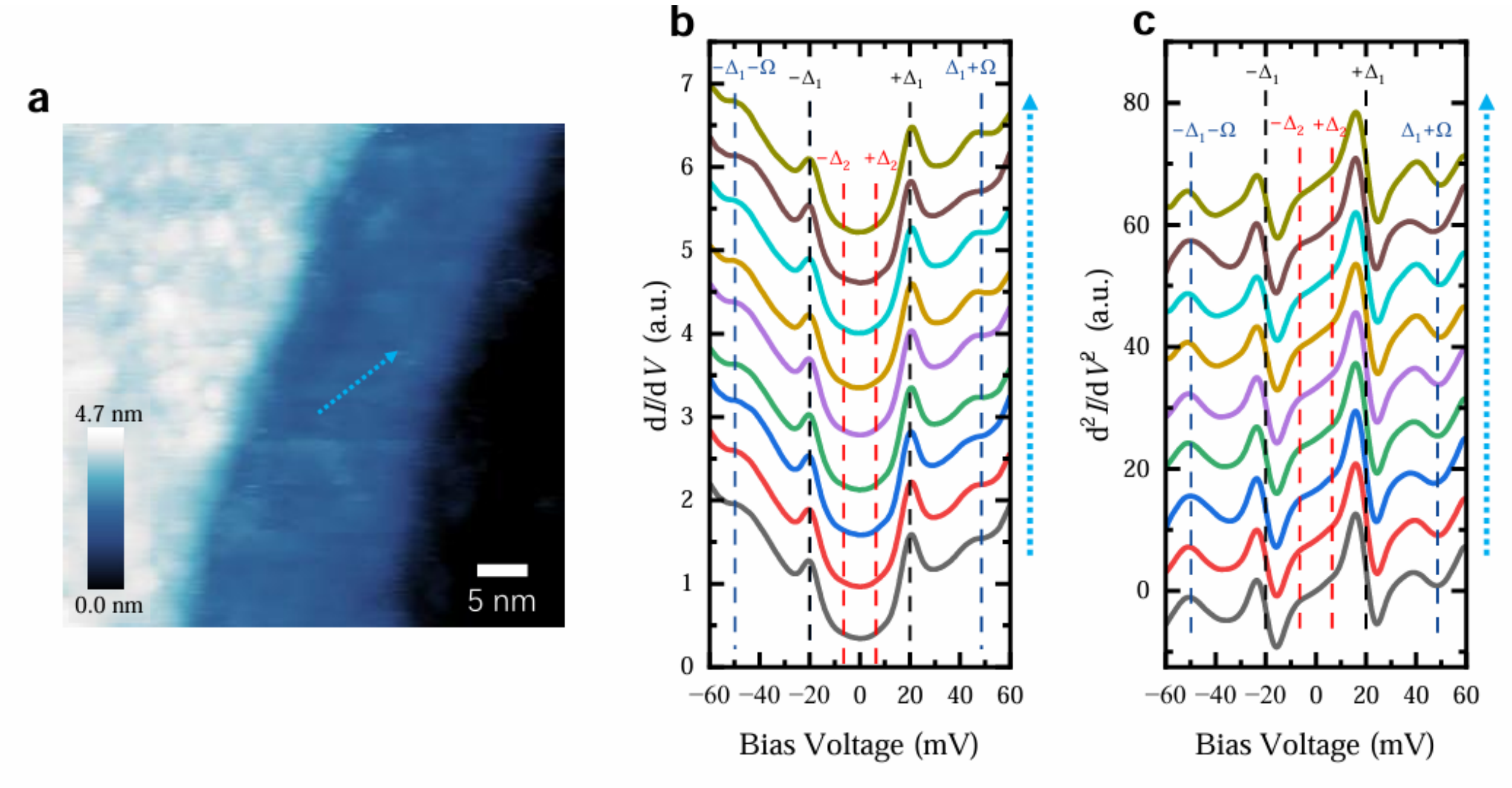

**Extended Data Fig. 4 | Tunneling spectra measured along another trajectory in the same area shown in Fig. 2. a**, Topographic image in the area as shown in Fig. 2, now the dotted arrowed line shows a different trajectory for measuring another set of spectra. **b**, A set of tunneling spectra measured along the dotted arrowed line in Layer 1 shown in **a**. The spectra look quite similar as those shown in Fig. 2. **c**, Both the two gap values and the bosonic mode energy scale can be obtained in the curves of $d^2I/dV^2$ versus $V$ corresponding to the spectra in **b**.

| Gap notation | $\Delta_1$ (meV) | $\Gamma_1$ (meV) | $\Delta_2$ (meV) | $\Gamma_2$ (meV) | $\Delta_1$ proportion |
|---|---|---|---|---|---|
| two anisotropic s-wave | 18.8 | 4.4 | 6 | 6 | 80% |
| anisotropic *s*-wave + *d*-wave | 18.8 | 4.1 | 6 | 6 | 80% |
| two *d*-wave | 20 | 2.2 | 6 | 5 | 80% |

**Extended Data Table I | The parameters used for the Dynes model fittings in Fig. 3.**

| Gap notation | $\Delta_1$ (meV) | $\Gamma_1$ (meV) | $\Delta_2$ (meV) | $\Gamma_2$ (meV) | $\Delta_1$ proportion |
|---|---|---|---|---|---|
| two anisotropic s-wave | 19 | 2.6 | 6 | 6.5 | 65% |
| anisotropic *s*-wave + *d*-wave | 19 | 2.9 | 5.5 | 5 | 70% |
| two *d*-wave | 19 | 2.8 | 6 | 4 | 80% |

**Extended Data Table II | The parameters used for the Dynes model fittings in the Extended Data Fig. 3 (in another region).**

# Supplementary Information for

# Superconducting gap structure and bosonic mode in $La_2PrNi_2O_7$ thin films at ambient pressure

Shengtai Fan[1†], Mengjun Ou[1†], Marius Scholten[2†], Qing Li[1], Zhiyuan Shang[1], Yi Wang[1], Jiasen Xu[1], Huan Yang[1*], Ilya Eremin[2*], Hai-Hu Wen[1*]

[1] National Laboratory of Solid State Microstructures and Department of Physics, Jiangsu Physical Science Research Center, Collaborative Innovation Center of Advanced Microstructures, Nanjing University, Nanjing 210093, China.

[2] Institut für Theoretische Physik III, Ruhr-Universität Bochum, D-44801 Bochum, Germany

*Corresponding authors: huanyang@nju.edu.cn, Ilya.Eremin@ruhr-uni-bochum.de, hhwen@nju.edu.cn

**Supplementary Note 1: Electronic structure in a wide energy window**

The tunneling spectra within a wide energy range from −1 V to 1 V (Supplementary Fig. 1a) were measured along the dashed arrowed line in Extended Data Fig. 2b. Notably, the spectra exhibit a gap-like line shape, with the significantly suppressed spectral weights around the Fermi level and the rapid increase of spectral weights on both sides at high energies. This gap-like line shape shares similarities to those observed in doped cuprates and other doped Mott insulators, indicating that the $La_2PrNi_2O_7$ thin film may have comparable electron correlations. Recent experiments[1,2] and corresponding theoretical calculations[3] have unveiled the presence of Mottness in $La_3Ni_2O_7$. Some experiments[4,5] and theoretical calculations[6,7] even demonstrate the presence of charge transfer characteristics in $La_3Ni_2O_7$. Here, we use black/pink dashed lines to indicate the linear fits to the spectra below and above the charge transfer band (CTB)/upper Hubbard band (UHB), and the crossing points are determined as the onset energy of CTB/UHB. The charge transfer energy $\Delta_{CT}$ could be defined as the difference between the onset energy of UHB and CTB. The statistical analysis (Supplementary Fig. 1b) suggests the $\Delta_{CT}$ value is approximately 1.51 ± 0.05 eV, which is close to some experimental findings[2,5] and theoretical predictions[7] in $La_3Ni_2O_7$. However, we are not completely sure that this gapped feature in high energy region is induced by the correlation effect or charge transfer band, or these two edges just correspond to the band edges of dense bands in those energy regions.

## Supplementary References

1. Liu, Z. *et al.* Electronic correlations and partial gap in the bilayer nickelate $La_3Ni_2O_7$. *Nat. Commun.* **15**, 7570 (2024).
2. Fan, S. et al. Tunneling spectra with gaplike features observed in nickelate $La_3Ni_2O_7$ at ambient pressure. *Phys. Rev. B* **110**, 134520 (2024).

3. Geisler, B., Hamlin, J. J., Stewart, G. R., Hennig, R. G. & Hirschfeld, P. J. Structural transitions, octahedral rotations, and electronic properties of $A_3Ni_2O_7$ rare-earth nickelates under high pressure. *npj Quan. Mater.* **9**, 38 (2024).

4. Dong, Z. *et al.* Visualization of oxygen vacancies and self-doped ligand holes in $La_3Ni_2O_{7-\delta}$. *Nature* **630**, 847-852 (2024).

5. Chen, X. *et al.* Electronic and magnetic excitations in $La_3Ni_2O_7$. *Nat. Commun.* **15**, 9597 (2024).

6. Wú, W., Luo, Z., Yao, D.-X. & Wang, M. Superexchange and charge transfer in the nickelate superconductor $La_3Ni_2O_7$ under pressure. *Sci. China Phys., Mechan. & Astron.* **67**, 117402 (2024).

7. Chen, X., Jiang, P., Li, J., Zhong, Z. & Lu, Y. Charge and spin instabilities in superconducting $La_3Ni_2O_7$. *Phys. Rev. B* **11,** 014515 (2025).

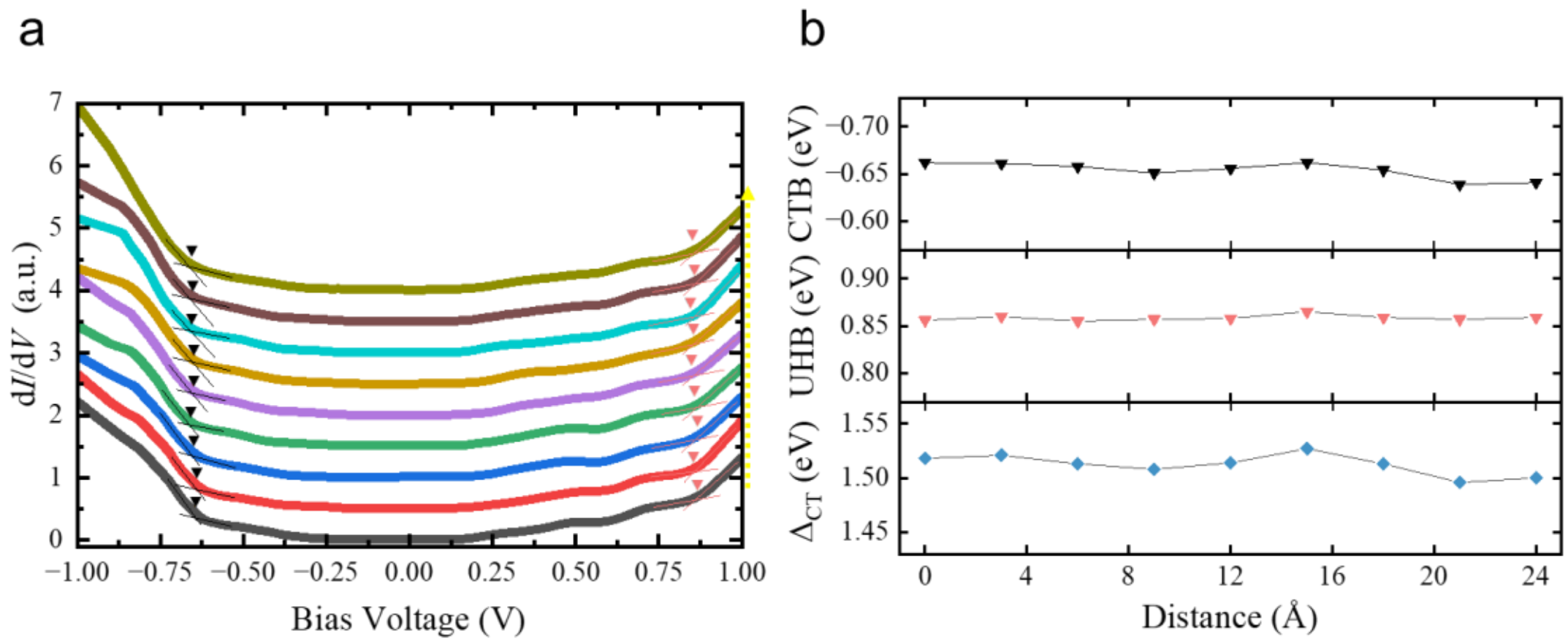

**Supplementary Fig. 1: Tunneling spectra within a wide energy range. a**, Tunneling spectra in the energy range of ±1.0 V measured along the dashed arrowed line in Layer 1, as shown in Extended Data Fig. 2b. The black/pink symbols in **b** indicate the energy values obtained by having linear fits to the spectra near the two gap edges, and the crossing points are determined as the onset energy of the supposed CTB/UHB. Black and pink triangles mark the onset positions of CTB and UHB. **b**, The onset energy of CTB (top panel), UHB (middle panel) and the charge transfer energy $\Delta_{CT}$ (bottom panel) as a function of the measured position along the dashed arrowed line in Extended Data Fig. 2b. (Setpoint conditions: $V_{set}$ = 1 V and $I_{set}$ = 100 pA).